\documentclass[11pt]{article}
\usepackage{blois}
\usepackage{verbatim}
\usepackage{tabularx}
\usepackage{graphicx}% Include figure files
\usepackage{dcolumn}% Align table columns on decimal point
\usepackage{amssymb}% bold math
\def\s{s_\beta}
\def\c{c_\beta}

\newcommand{\eq}[1]{Eq.~\ref{#1}}
\newcommand{\tev}{\, {\rm TeV}}
\def\beq{\begin{equation}}
\def\eeq{\end{equation}}

\def\be{\begin{equation}}
\def\ee{\end{equation}}
\def\bea{\begin{eqnarray}}
\def\eea{\end{eqnarray}}

\newcommand{\Photo}{\includegraphics[height=35mm]{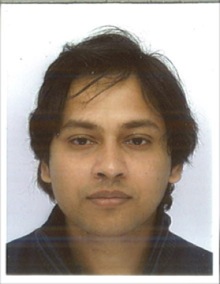}}

\begin{document}
\vspace*{4cm}
\title{BSM THEORIES FACE HIGGS COUPLING DATA}

\author{ RICK S GUPTA }

\address{IFAE, Universitat Autonoma de Barcelona, 08193 Bellaterra, Barcelona, Spain}

\maketitle\abstracts{
We  discuss how much Higgs couplings (including the Higgs self coupling)  can deviate from their Standard Model values, in different Beyond Standard Model (BSM) theories, if no other BSM states are accessible at the LHC.  Then, we  focus on supersymmetric theories and show that there is a connection between  the mechanism to raise the Higgs mass  and the pattern of Higgs coupling deviations. }

\section{Introduction}

{ If there is new physics beyond the Standard Model (BSM) that stabilizes the Higgs mass,  it will  lead to deviations in its couplings.  Just as LEP probed scales ($\sim$ 3 TeV) much higher than its center of mass energy (209 GeV), the hope is that there are instances where Higgs coupling measurements can similarly
surpass direct searches. Higgs coupling measurements become crucial  precisely in the regions where direct searches are difficult, because,  in the absence of direct search discoveries, Higgs coupling deviations would  become primary evidence for an exotic Higgs sector. The maximum allowed deviation in Higgs couplings  with respect to the SM such that no BSM state is accessible at the LHC, even in the long run, should thus serve as a target for the measurement precision of Higgs couplings.~\cite{Gupta1,Gupta2}
One of the main objectives of the paper is to find these targets in different  BSM examples. This is discussed in Sec.~Ê2. In Sec.~3  we will show in that in supersymmetric models, a precise relationship exists between the mechanism to raise the tree level Higgs mass and the pattern of coupling deviations.~\cite{Gupta3} 
\section{How well do we need to measure Higgs boson couplings?}
\label{target}

\begin{table}[t]
\label{targetT}
\caption{\label{table:results} Targets for Higgs couplings in different BSM theories.  In the last row we show projected $1\sigma$ LHC sensitivities at $14\tev$ with  $3\, {\rm ab}^{-1}$ data, based on the results of Klute et al.}
\begin{center}
\begin{tabular}{|lcccc|}
\hline
       & $\Delta hVV$ & $\Delta h\bar tt$ & $\Delta h\bar bb$& $\Delta hhh$ \\
\hline
Mixed-in Singlets & 6\% & 6\% & 6\% & -18\%\\
Composite Higgs & 8\% & tens of \% & tens of \%& tens of \% \\
MSSM & $<1\%$ & 3\% &  $100\%$&15$ \%$ \\ %\hline
LHC $14\tev$, $3\, {\rm ab}^{-1}$~\cite{Klute:2012pu}&  $8\%$ & $10\%$ & $15\%$& -30 \%, +20 \% \\
\hline
\end{tabular}
\end{center}
\vspace{-0.6cm}
\end{table}

\subsection{Mixed in Singlets}

 In models with new singlet states there are two Higgs boson mass eigenstates, $h$ and $H$, due to mixing between  the singlet and SM neutral CP even states, $h$ being the SM-like Higgs boson. For Higgs couplings to vectors and fermions, these states share the SM value of the Higgs coupling squared: $g_h^2 =c_h^2~g^2_{{SM}},~~g^2_H= s_h^2~g^2_{{SM}}$, where $s_h=\sin \theta_h$ and $c_h= \cos \theta_h$, and $\theta_h$ is the mixing angle between the gauge eigenstates. On the other hand, using the potential in~\cite{Bowen} we find that the Higgs self coupling is given by $
 g_{hhh}=g_{hhh}^{SM}(c_h^3-1)$. Fig.~1(left) shows the upper bound on $s_h^2$ from precision data. While we want to stay in the region allowed by precision data, we also want that the mass of the exotic Higgs $H$, $m_H$, is large enough so that $H$ is barely inaccessible at the LHC. Fig.~1(left)  also shows a detectability curve for the exotic state $H$, adapted from the study by Bowen et al,~\cite{Bowen}  which studies the prospect of observing the state $H$ with 100 fb$^{-1}$ data at the LHC. It is clear from Fig.~1(left) that the maximum value of $s_h^2$ consistent with precision data but for which $H$ is inaccessible at the LHC is given by the intersection of the two curves. This gives: $ {s_h^{2}}_{max}  \approx 12 \%,~~~ (\Delta g_{h}/g_{{SM}})^{target}\approx -6 \%,~~~
\Delta g^{target}_{hhh}/g^{SM}_{hhh}= -18 \%$.

\subsection{Composite Higgs models}

In composite models Higgs coupling deviations of ${\cal O}(\xi)$ are expected, where $\xi=v^2/f^2$ and $f$ is the `pion decay constant' of the strongly coupled theory.  The strongest constraint on $\xi$ comes from the  precision measurements: $\xi \lesssim 0.15$ at 90 $\%$ CL. Direct LHC probes are expected to be much less sensitive than existing precision constraints.~\cite{clic} The precision upper bound $ \xi \sim 0.15$  sets the target for composite Higgs coupling deviations to be of the order of  tens of $\%$.

\subsection{The Minimally Supersymmetric Standard Model}

Now we turn to Higgs couplings deviations in  the MSSM. As we will explain in detail in the next section deviations in the Higgs couplings to up-type quarks and vector bosons are suppressed in the MSSM. We will therefore present results on the self coupling and the down-type couplings. Fig.~1(right) shows the results of a scan performed using  {\tt FeynHiggs2.8.6}~\cite{FeynHiggs} for the $hbb$ coupling.   The points in red represent points where more than one Higgs boson can be seen with 300 fb$^{-1}$ LHC data,  according to ATLAS projections.~Ê\cite{tdr} To find the target, we have to find the largest  deviation, excluding the red  points in Fig.~1(right). We find that large measurable deviations are possible for low tan beta ($\tan \beta \sim 5$). This is because  in this region of the parameter space  even very small values of $m_A \sim 200$ GeV are inaccessible with 300 fb$^{-1}$ LHC data and as we will show in the next section small $m_A$ values correspond to large Higgs coupling deviations.  For the Higgs self coupling a similar analysis~\cite{Gupta2} shows that appreciable deviations can again occur only for  $\tan \beta\sim 5$. For the self coupling in the MSSM we find the target value of -18$\%$.~\cite{Gupta2} In Table~1 we summarize the Higgs targets for the three classes of models we have considered. We also present the anticipated LHC sensitivity to Higgs coupling deviations~\cite{Klute:2012pu} which show that these targets are beyond LHC capability. This makes a very strong case for linear colliders which have greater measurement precision. 
 %If you more commonly use the method of square brackets in the line of text
%for citation than the superscript method,
%please note that you need  to adjust the punctuation
%so that the citation command appears after the punctuation mark.
\begin{figure}[t]
\label{hbb}
\centering
\begin{tabular}{cc}
\includegraphics[width=0.38\columnwidth]{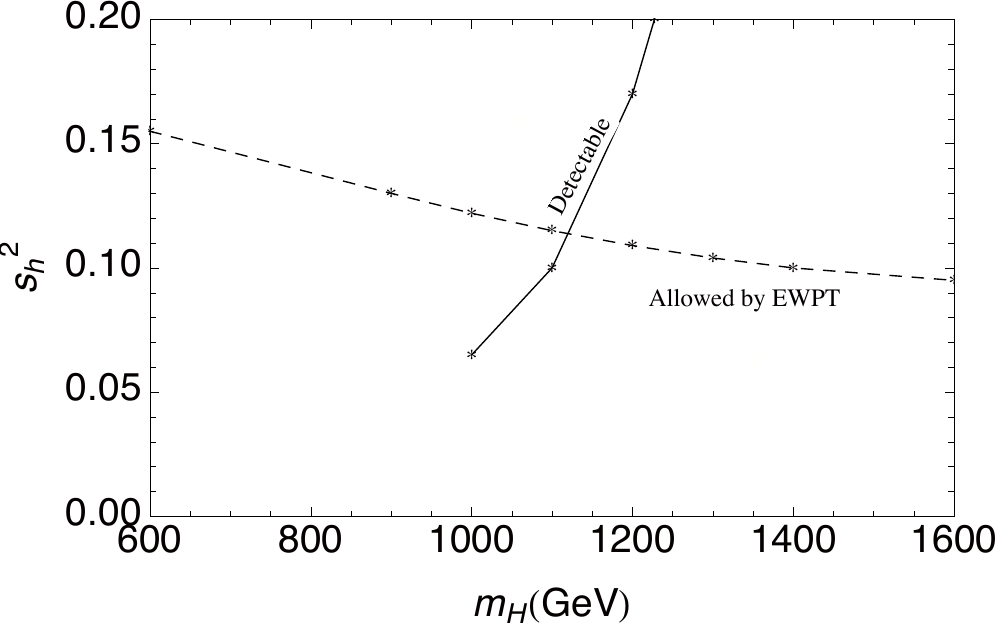} &
\includegraphics[width=0.38\columnwidth]{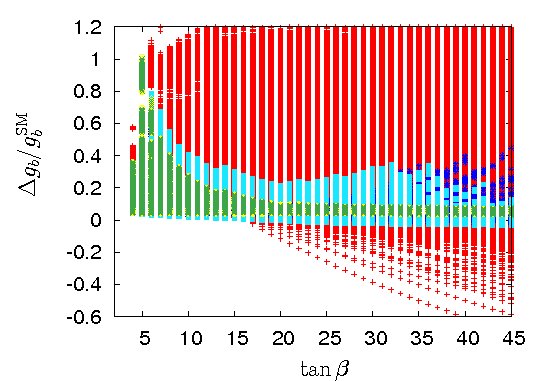} \\

\end{tabular}

\caption{Left: Electroweak precision (dashed) and detectability (solid) constraints on $s_h^2$. Right: We show $\Delta g_b/g_b^{SM}$ 
as a function of $\tan \beta$. The colour coding is as follows: red means more than one Higgs boson can be discovered at the LHC - for all the other colours
  a single Higgs boson discovery would be discovered at the LHC.  }
\end{figure}
\begin{figure}[t]
\centering
\begin{tabular}{cc}
\includegraphics[width=0.25\columnwidth]{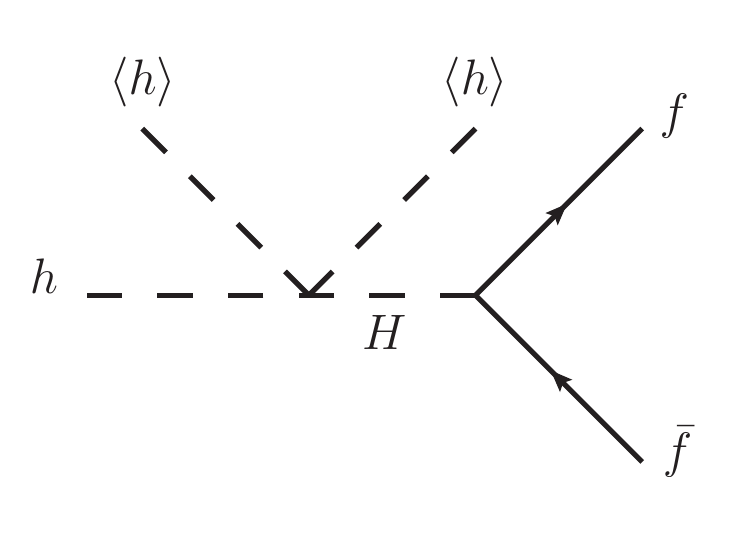} &
\includegraphics[width=0.25\columnwidth]{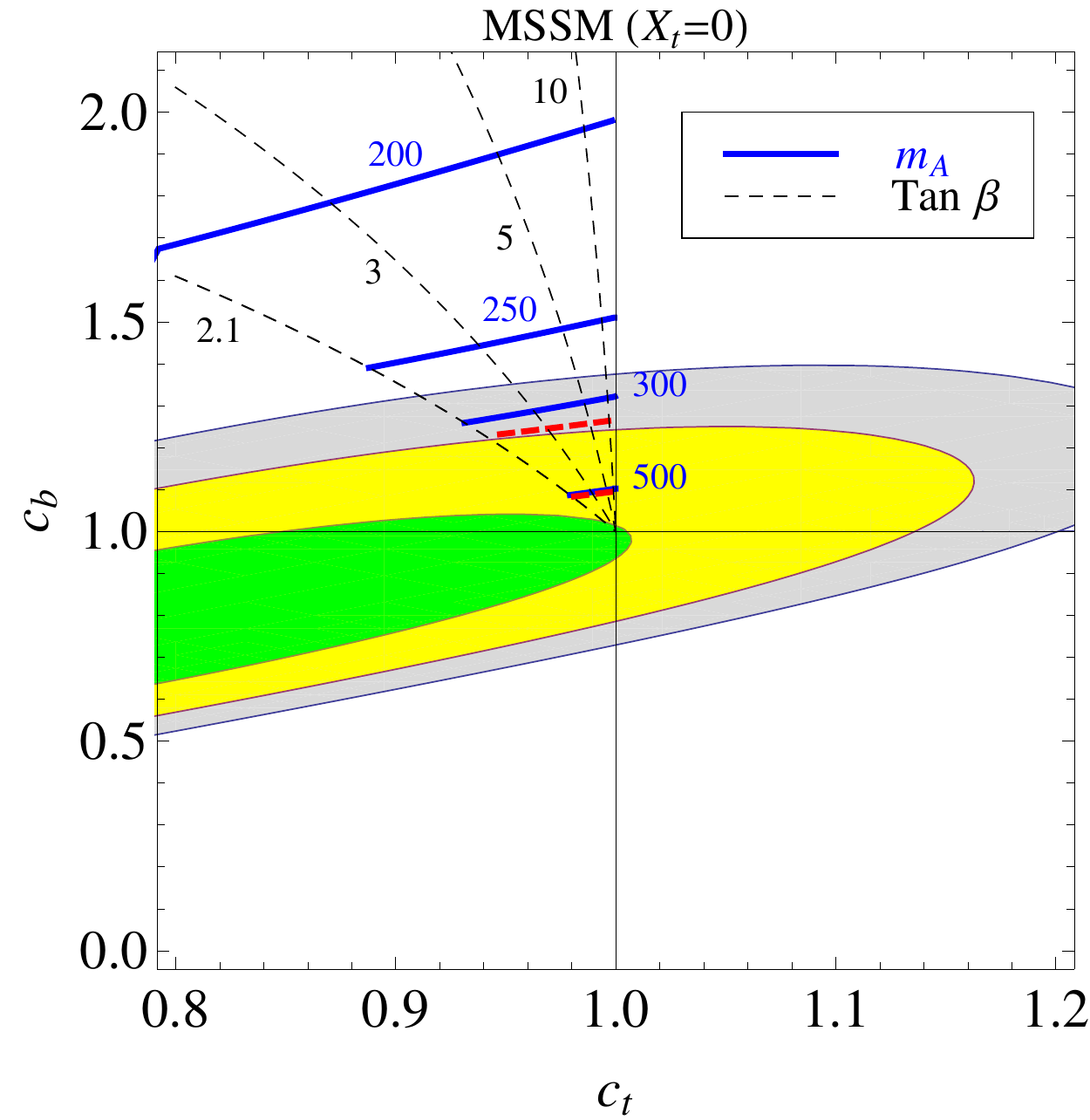} \\
\includegraphics[width=0.25\columnwidth]{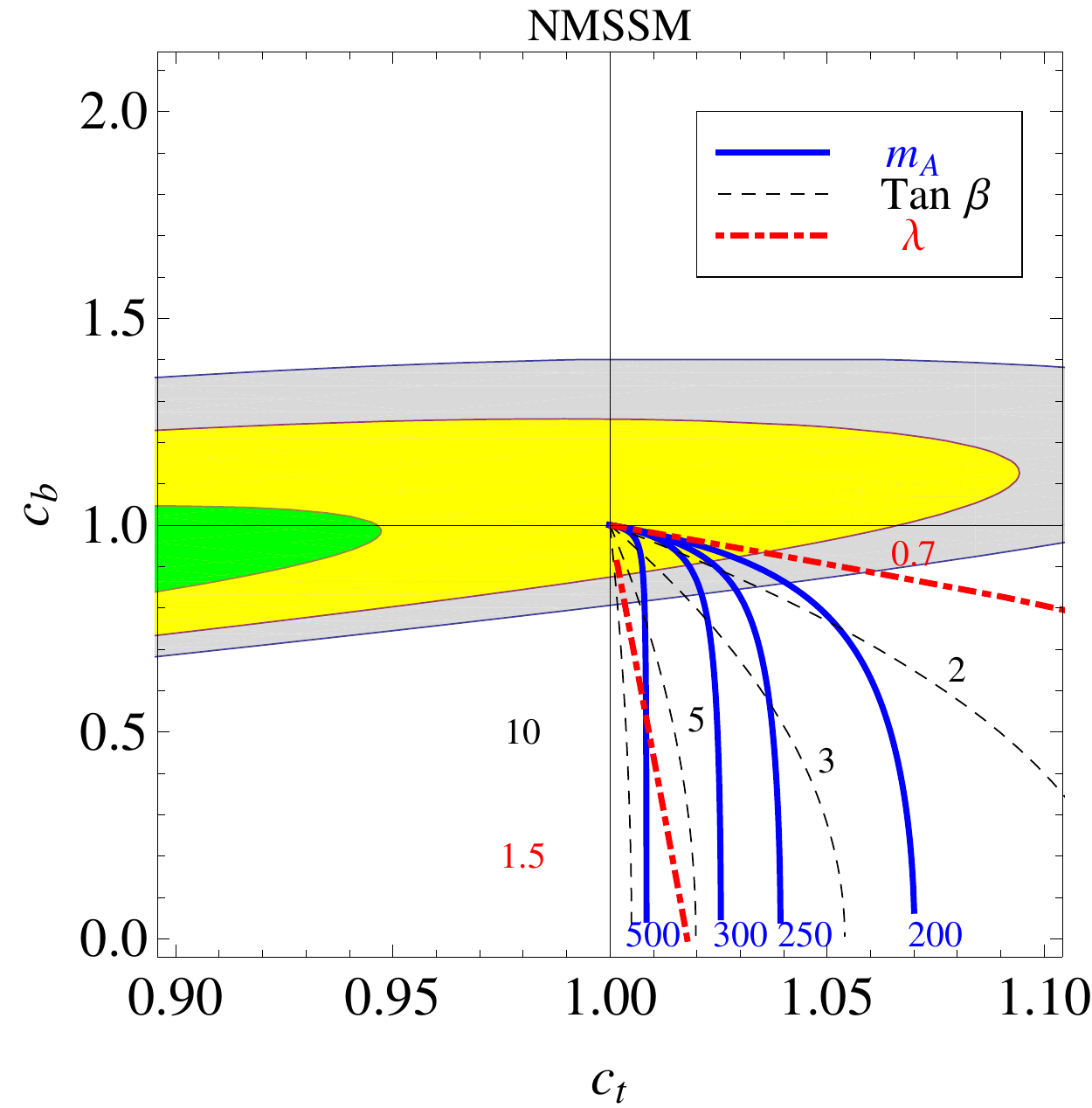} &
\includegraphics[width=0.4\columnwidth]{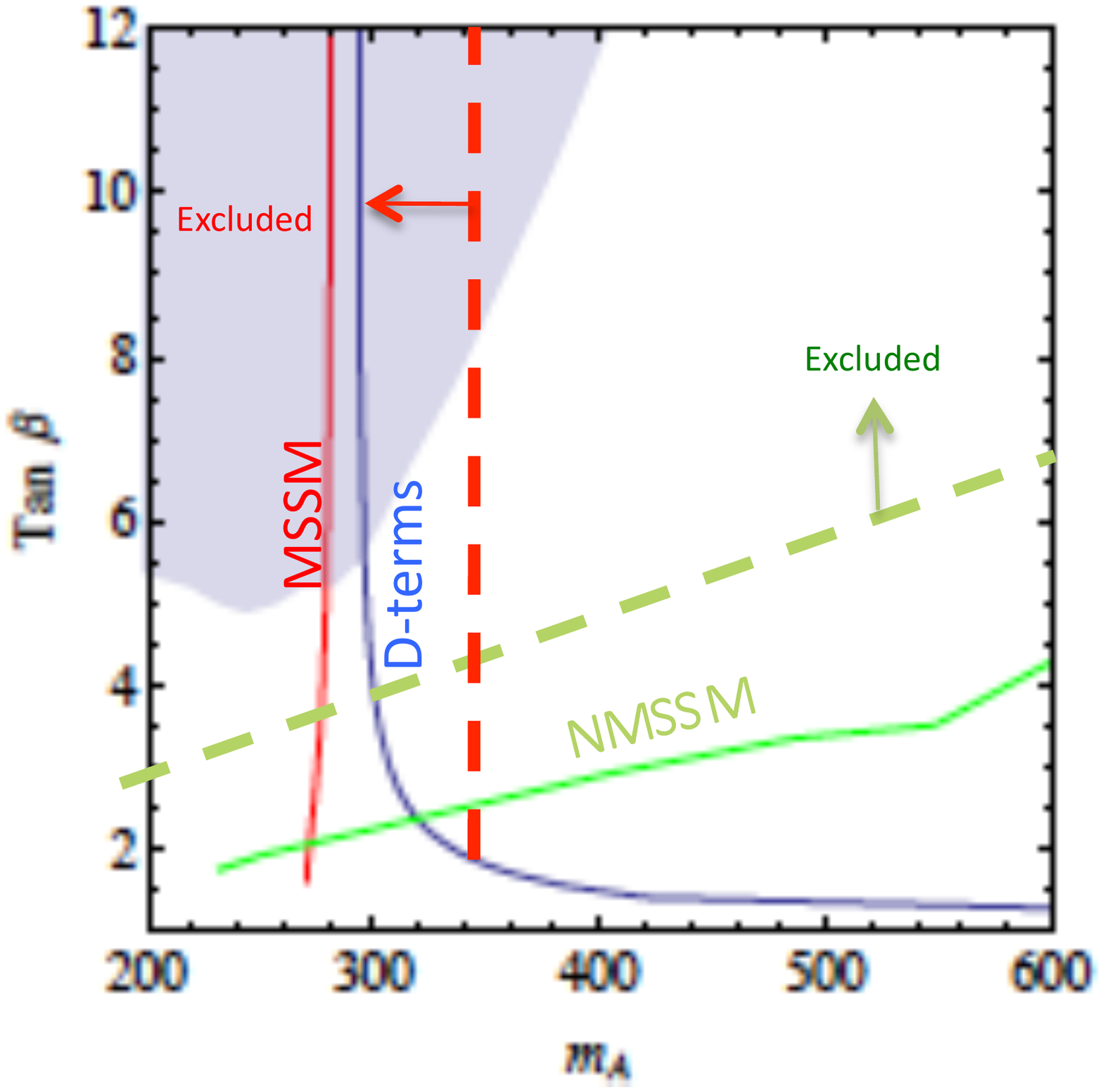}
\end{tabular}
\caption{In the top left we show the Feynman diagram responsible for causing deviations in the the SUSY Higgs couplings from their SM values. We also show Higgs coupling deviations in the MSSM with heavy stops and no mixing (top right) and the NMSSM (bottom left). For the MSSM we use heavy stops (assuming no mixing between them) to obtain the 125 GeV Higgs. For the NMSSM we assume the existence of 500 GeV stops and then fix $\lambda$ by requiring the correct Higgs mass.  We also show the 68 $\%$ (green), 95 $\%$ (yellow) and 99 $\%$ (grey) C.L. regions obtained by performing a global fit over LHC data. In the bottom right plot we show the 95 $\%$ CL exclusion curves for the different models. The grey region shows the bound from CMS $H \to \tau \tau$. We also show exclusion curves (dashed lines) obtained by Barbieri et al, using similar methods, but more recent data.}
\label{bodo}
\end{figure}

\section{The Higgs Mass/Couplings Connection in SUSY} \label{secsusy}

In SUSY the connection of  the Higgs potential of the doublets, $H_{1,2}$ to the mass and couplings of the lightest Higgs boson is clearest in the basis  where only one of the Higgs doublets gets a vacuum expectation value (VEV). We can transform to this basis, $\{h,H\}$,  by carrying out  a rotation of the neutral CP even gauge eigenstates ${h_1^0,h_2^0}$  by an angle $\beta =$ArcTan[$v_u/v_d$]. In the limit that the contribution of the  quartic couplings, $\delta_i$, to the mass matrix can be neglected in comparison to the quadratic terms which are  ${\cal O}(m_H^2)$ (the so called decoupling limit where, $\delta_i v^2/m_H^2 \ll 1$), this basis is also the mass eigenstate basis, $H$ being the heavier Higgs. As the neutral CP even part of one of  the doublets gets the full VEV, $v= 246$ GeV,  the physical Higgs from this doublet, $\tilde{h}= (h-v)$, has exactly SM couplings. As $\delta_i v^2/m_H^2 $ increases, the misalignment between the mass basis and the  $\{h,H\}$ basis  cannot be neglected and the lightest Higgs boson mass eigenstate shows deviations from SM Higgs coupling values.  Consider a general quartic contribution to the potential expressed in the $\{h,H\}$ basis,
\begin{equation}\label{effl}
\Delta V(H_1,H_2)= \delta_\lambda h^4+\delta h^3 H+
\delta_2 h^2 H^2+\delta_3 h H^3+\delta_4 H^4,
\end{equation}
The $\delta_\lambda h^4$ contributes to the mass of the lightest Higgs $\Delta m_h^2 =8 \delta_\lambda v^2$. The $\delta h^3 H$ term causes a mixing between $h$ and $H$ states which leads to new contributions to $hff$ couplings via the diagram in  Fig.~2 (top left). Note that the other terms proportional to $\delta_{2,3,4}$ do not lead to a mixing between $h$ and $H$ because at the lowest  order only $h$ gets a VEV. 

Taking into account the contribution of Fig.~2 (top left)  and the fact that the same diagram also modifies the fermion mass we finally get,
\begin{equation}
c_f\equiv \frac{y_{f}}{\sqrt{2}m_f/v}\approx\frac{Y_f^h- 3Y_f^H \delta \frac{v^2}{m_H^2}}{Y_f^h- Y_f^H \delta \frac{v^2}{m_H^2}}\approx 1-2\delta \frac{Y_f^H}{Y_f^h}  \frac{v^2}{m_H^2}\, .
\label{cbct}
\end{equation}
where $Y^H_f$ is the coupling of $H$ to fermions $f=t,b,\tau$ and $Y_f^H/Y_f^h= \tan \beta (-\cot \beta)$ for down-type  (up-type) quarks. Thus any modification of the Higgs potential, via the $\delta_\lambda$ coupling, is generally accompanied by a contribution to the $\delta$ coupling, which modifies Higgs couplings in a correlated way. For Higgs coupling to vectors there is no diagram that gives an ${\cal O}(\delta v^2/m_H^2)$ contribution and the lowest order contribution is ${\cal O}(\delta^2 v^4/m_H^4)$.  In Table 2 we present the expressions for $\Delta V, \delta_\lambda$ and $\delta$  for the MSSM at tree level and  the NMSSM (with the singlet decoupled). Because of the difference in sign in $\delta$ the theoretical deviations go in different  directions (top left/bottom right) in the $c_b-c_t$ plane for the MSSM/NMSSM in accordance with \eq{cbct}. We show the theoretical Higgs coupling  deviations and the corresponding exclusion curves in the $m_A-\tan \beta$  for these two models in Fig.~\ref{bodo}.  An analysis with more recent data has been performed by Barbieri et al;~\cite{barbieri} we show their exclusion curves by dashed lines. The expressions for $\delta_\lambda$ and $\delta$ for other mechanisms to raise the Higgs mass, like large A-terms, additional D-terms etc can be similarly derived.~\cite{Gupta3}   
\begin{table}[t]
\caption{\label{cbct} The expressions for $\Delta V, \delta_\lambda$ and $\delta$ in the MSSM and NMSSM. }
\begin{center}
\begin{tabular}{|lccc|}
\hline
       & $\Delta V$ & $\delta_\lambda$ & $\delta$\\
\hline
MSSM & $\frac{g^{2}+g^{\prime\,2}}{32}\left((h_1^0)^2-(h_2^0)^2\right)^2$& $\frac{m_Z^2}{16 v^2}(\c^2-\s^2 )^2$ &$\frac{m_Z^2}{ 2v^2} \s \c (\c^2-\s^2)$\\&&&Ê\\

NMSSM& $\frac{\lambda^2}{4} (h^0_1 h^0_2)^2$ & $\frac{\lambda^2}{16}\sin^2 2 \beta$&$-\frac{\lambda^2}{8}\sin 4 \beta$\\ %\hline
\hline
\end{tabular}
%~\cite{Klute:2012pu}.}
\end{center}
\vspace{-0.6cm}
\end{table}

\newpage
\section*{Acknowledgments}

This paper is based on two earlier works. So first of all I would like to thank my collaborators, M. Montull, F. Riva, H. Rzehak and J.D. Wells. I would also like to thank the organizers of the conference, especially C. Grojean, for inviting me.

\end{document}